\input harvmac

\def\npb#1(#2)#3{{ Nucl. Phys. }{B#1} (#2) #3}
\def\plb#1(#2)#3{{ Phys. Lett. }{#1B} (#2) #3}
\def\pla#1(#2)#3{{ Phys. Lett. }{#1A} (#2) #3}
\def\prl#1(#2)#3{{ Phys. Rev. Lett. }{#1} (#2) #3}
\def\mpla#1(#2)#3{{ Mod. Phys. Lett. }{A#1} (#2) #3}
\def\ijmpa#1(#2)#3{{ Int. J. Mod. Phys. }{A#1} (#2) #3}
\def\cmp#1(#2)#3{{ Commun. Math. Phys. }{#1} (#2) #3}
\def\cqg#1(#2)#3{{ Class. Quantum Grav. }{#1} (#2) #3}
\def\jmp#1(#2)#3{{ J. Math. Phys. }{#1} (#2) #3}
\def\anp#1(#2)#3{{ Ann. Phys. }{#1} (#2) #3}
\def\prd#1(#2)#3{{ Phys. Rev.} {D\bf{#1}} (#2) #3}
\Title{hep-th/9606086, HUTP-96/A021,OSU-M-96-10
}{\vbox{\centerline{Matter From Geometry}}}
\centerline{Sheldon Katz}
 \medskip\centerline{\it Department of Mathematics}
 \centerline{\it Oklahoma State University}
 \centerline{\it Stillwater, OK 74078, USA}
\vskip 0.15in
\centerline{and}
\centerline{Cumrun Vafa }
\medskip\centerline{\it Lyman Laboratory of Physics}
\centerline{\it Harvard University}\centerline{\it Cambridge, MA
02138, USA}
\vskip 0.15in
\vskip .3in

We provide a local geometric
description of how charged matter arises in type IIA, M-theory, or
F-theory compactifications on Calabi-Yau manifolds.
The basic idea is to deform a higher singularity into a lower
one through Cartan deformations which vary over space.  The results
agree with expectations based on string dualities.
\Date{6/96}

\newsec{Introduction}
In a recent work \ref\sau{M.~Bershadsky, K.~Intriligator,
S.~Kachru, D.R.~Morrison, V.~Sadov, C.~Vafa, {\it Geometric
Singularities
and Enhanced Gauge Symmetries\/}, hep-th/9605200.}\ a
detailed check was made of the enhanced gauge symmetry points of
F-theory/heterotic duality
in $d=6$ with $N=1$ supersymmetry proposed in
\ref\mv{D.R.~Morrison and C.~Vafa, {\it Compactifications of F-Theory
on
Calabi--Yau Threefolds---I,II\/}, hep-th/9602114,9603161.} .  There
it was shown how charged matter should arise
upon compactification of F-theory to a Calabi-Yau threefold
for consistency with F-theory/heterotic duality.
As far as the matter content is concerned various results were found
in \sau\ based on the duality with heterotic strings \mv\
but a purely F-theory (type IIA) derivation for the matter content
was lacking.
In particular it was observed there that for
simply laced gauge groups very often the matter seems localized at
`extra'
singularities of the manifold.
  Here we wish to derive this structure from the viewpoint of
F-theory alone thus sharpening the F-theory/heterotic
duality check.
 A heterotic derivation of this localization
has been recently done by Witten \ref\wit{E. Witten, to appear.}.
In fact the physical interpretation of the localization
we find is identical to that found there.

We will recover not only the matter
structure/geometry dictionary anticipated in \sau\
but also present examples of new such cases. Our aim here is not
to be exhaustive in this matter/geometry dictionary but only to
present
the main idea through examples.

\newsec{Basic Idea}

We  first consider compactification of F-theory to 8 dimensions (or
type IIA
to 6 dimensions)
on an elliptic (general) $K3$ \ref\cv{C. Vafa, {\it Evidence for
F-theory\/},
HUTP-96/A004, hep-th/9602022.}.
  Suppose we are at a moduli value of $K3$
where we have a singularity of A-D-E type.
In this case the gauge symmetry is the corresponding
A-D-E group.
We consider a further compactification of this theory to lower
dimensions.
For concreteness let us consider further
compactification on a one dimensional complex space
denoted by a parameter $t$, where the
moduli of $K3$ is varying over this space.  We wish to find
the gauge group and matter representation in the lower dimensional
theory.  The gauge group is easy to identify simply by considering
what is the singularity type over generic $t$ \ref\gra{P.~Aspinwall
and
M.~Gross, {\it The $SO(32)$ Heterotic String on a K3 surface\/},
CLNS-96/1409,hep-th/9605131.}\sau ;
 there could also be further
monodromy acting on the singularity leading to non-simply
laced groups \gra .  Let us concentrate on the case where there
are no further monodromies, i.e.\ what is called the `split' case
though our results can be partially
generalized to the other case as well.

To fix our terminology, let $G$ denote the corresponding
A-D-E group.  There is a complex scalar field in the adjoint
representation of $G$. For a generic expectation value
of this scalar the group $G$ is broken to $U(1)^n$ where $n$
is the rank of $G$.  For special $U(1)$'s we can have
$H\times U(1)\subset G$ where $H$ has no
$U(1)$ factors and has rank one lower than that of $G$.
The basic idea is to consider fiberings where the scalar field
in this $U(1)$ direction is identified with the
fibration parameter $t$.  Thus the surviving
gauge symmetry in the lower dimensional theory
 is $H$.  The massless matter representation for $H$ can
be read off by decomposing the matter which is in the
adjoint representation of $G$ in terms of $H\times
U(1)$
representations and by finding the zero modes of the Dirac operator
coupled to the gauge bundle given by $U(1)$.  Let $(R^a, q^a)$ denote
the representations we get where $q^a$
 denote the $U(1)$ charge.  If the fibration were trivial
the gauge symmetry would have been $G$ and the matter would
have been in the adjoint representation of $G$.  This is still
true at one point on the fibration $t=0$ where $G$ symmetry is
restored.  So we still will get the adjoint representation of $G$,
but the effect of fibration is to make that matter a representation of
$H$
according to how the adjoint decomposes.
To see this let $t$ denote the complexified Cartan of the $U(1)$,
which
we identify with the fibering parameter.
Then the Dirac operator giving the number of zero modes
of $R^a$ is given by solving
\eqn\dir{[D +q_a t]\psi_a (t,\bar t)=0}
where $D$ denotes the Dirac operator on $t$--plane.
The number of $R^a$ representations we will get is given by
the number of normalizable
zero modes of \dir\ which in turn is one
(if $q_a\not= 0$)--note that
the zero mode is localized near $t=0$ which
 after a suitable rescaling of $t$ goes as
 $\propto {\rm exp}(-t\overline t)$)\foot{ This idea
is well known in the context of family's index theorem applied
to the case at hand where $t$ is the
parameter space and the localization of the zero modes occurs
at some points on the parameter space.}.
Here we are concentrating on the matter localized at the `extra'
singularities
and so our considerations are all local
with respect to the $t$--parameter; we are not concerned with other
matter which are not concentrated at
these extra singularities\foot{
This is why the representations with $q_a=0$ are not relevant for our
considerations.}.  There are cases where the matter is not
localized in this way.  For example,  in the type IIA theory
when we have a genus $g$ curve of singularities of $A_n$ type
we get $g$ adjoints of $A_n$ \ref\kmp{S. Katz, D. Morrison and
R. Plesser, {\it Enhanced Gauge Symmetry in Type II String Theory},
hep-th/9601108.}\
which are not localized on any specific points on the curve.

As far as the description of the matter is concerned we can
give two alternative descriptions, depending on whether
we are talking about F-theory down to six dimensions or type IIA
down to 4. For F-theory description, let us assume that $G=SU(n)$,
where the geometry is realized by $n$ coinciding 7-branes.  Then
as a function of $t$ the location of the 7-branes is changing
and the resulting open strings pick up mass as a function of $t$ and
we are finding the wave function of the massless modes
of open strings concentrated near $t=0$.  As far as the type IIA
description is concerned, we have an adjoint of $G$ worth of
vanishing 2-cycles, with D-branes wrapped around them.
As we move away from $t=0$ some of the 2-cycles pick up mass
and the wave function of the 2-branes are concentrated near $t=0$.

Note that if we replace $t\rightarrow P_n(t)$, where $P_n(t)$
is a polynomial of degree $n$ in $t$, since the above
considerations are local, we learn that we obtain $n$ times
the matter we get, localized near the zeroes of $P_n(t)$.

To use the above basic idea what one needs to know is
how the expectation value of the scalar taking its
value in the complexified
 Cartan space ${\cal C}$ of the A-D-E modifies and resolves
the singularity.  Luckily this is mathematically known.

\newsec{Cartan Resolution of Singularities}
As discussed above one needs to know how
giving expectation values to the Cartan elements of the singularity
deform (and resolve) the singularity.  This is a well known
mathematical result.  A unified treatment is given
in~\ref\km{S.~Katz and D.R.~Morrison,
Jour.\ Alg.\ Geom.\ {\bf 1} (1992) 449.}, which contains references
to earlier foundational works.  The deformation
space of the resolved singularity can be identified
with the Cartan subalgebra.  Here we describe the result, which gives
the conditions
on these expectation values for certain curves in the resolution to
deform.
We do this separately for $A_n,D_n,E_n$.  In each case, we let
$\{e_i\}$
denote an orthonormal set of vectors.

The $A_n$ Cartan subalgebra $h$
is the space of vectors $\sum_{i=1}^{n+1} t_ie_i$
subject to the constraint $\sum t_i=0$.
The deformation space of the
resolution of the $A_n$ singularity similarly has coordinates
$(t_1,\ldots,t_{n+1})$
subject to $\sum t_i=0$.

The roots are in the dual space $h^*$, and are all of the form
$e_i^*-e_j^*$ for $i\neq j$.  The simple roots are
$v_i=e_i^*-e_{i+1}^*$.
Each $v_i$ corresponds to a vertex of the Dynkin diagram and
to a curve $C_i$ in the resolution of the $A_n$ singularity.  To each
positive root $e_i^*-e_j^*$ (so that $i<j$),
there is associated a set of vanishing 2-cycles
$C_{ij}=C_i+\ldots+C_{j-1}$,
and the condition on the $t_k$ for $C_{ij}$ to remain a vanishing
2-cycle is that
$t_i-t_j=0$.
The deformation of the $A_n$ singularity when we deform by
expectation
values in the Cartan is given by
\eqn\anc{xy+\prod_{j=1}^{n+1}(z+t_j)=0.}

The $D_n$ Cartan subalgebra is the space of vectors $\sum_{i=1}^n
t_ie_i$.
The deformation space of the resolution of the $D_n$ singularity has
coordinates $t_1,\ldots,t_n$.
The simple roots are $v_i=e_i^*-e_{i+1}^*$ for $1\le i\le n-1$, and
$v_n=
e_{n-1}^*+e_n^*$.  Each $v_i$ corresponds to a vertex of the Dynkin
diagram and
to a curve $C_i$ in the resolution of the $D_n$ singularity.  The
positive
roots are all of the form $e_i^*\pm e_j^*$.  In terms of the $v_i$,
the positive
roots are of one of the five forms $v_i+\ldots+v_{j-1}$, $v_n$,
$v_j+\ldots+v_{n-2}+v_n$, $v_j+\ldots+v_n$,
$v_j+\ldots+v_{k-1}+2v_k+\ldots
+2v_{n-2}+v_{n-1}+v_n$.  To each positive root $r$ is associated a
curve
$C$ by substituting $C_i$ for $v_i$ in the above expressions of the
root.
The $D_n$ singularity deformed by the Cartan parameters is given as
\eqn\dcar{x^2+y^2z-{\prod_{i=1}^n(z+t_i^2)-\prod_{i=1}^nt_i^2\over
z}+
2y\prod_it_i=0.}

The $E_n$ Cartan subalgebra is the space of vectors $\sum_{i=0}^n
a_ie_i$
subject to the constraint $-3a_0+\sum_{i=1}^na_i=0$.
The deformation space of the resolution of the $E_n$ singularity can
be
assigned coordinates $t_1,\ldots,t_n$, where $t_i=a_0/3+a_i$.
The simple roots are $v_0=e_0^*-e_1^*-e_2^*-e_3^*$,
and $v_i=e_i^*-e_{i+1}^*$ for $1\le i\le n-1$.  Each $v_i$
corresponds to a vertex
of the Dynkin diagram and
to a curve $C_i$ in the resolution of the $E_n$ singularity.
The deformation of $E_n$ singularities in terms of Cartan
parameters is more complicated.  The coefficients of the $E_6$ and
$E_7$
deformed polynomials are given in~\km, Appendices~1 and~2.
The $E_8$ case is known implicitly \km .

\newsec{The Cases}
As mentioned in the introduction the aim is to illustrate
how local geometric singularity encodes charged matter, and not so
much
to provide an exhaustive matter/geometry dictionary.  In this spirit
we provide some illustrative examples below.  It is possible
to check in all the cases below that have an overlap with results
based on F-theory/heterotic duality studied in \sau, the local
structure of the singularity is in agreement with expectations
(there is a change of notation relative to \sau\ where
there roughly is an interchange in what we denote as $x,y$).
However, in some cases (with real representations) we obtain
one hypermultiplet corresponding to appearance of the deformation
by $t^2$ in the singularity.  This is consistent with the
results of \sau\ where the appearance of $t$ in the singularity
was associated with one half a hypermultiplet in the
real representation.

\subsec{$A_n\to A_{n-k}\times A_{k-1}$}
We consider the breaking
$$SU(k)\times SU(n-k+1) \times U(1) \subset SU(n+1)$$
In this case by the decomposition of the adjoint of $A_n$
we expect the charged matter to be in the $({\bf k},{\bf n-k+1})$
representation of $SU(k)\times SU(n-k+1)$.

The above breaking of $A_n$
can be done by choosing the $U(1)$ Cartan to correspond
to the $t$ direction given by
$$t_1-t_2=\ldots=t_{k-1}-t_{k}=t_{k+1}-t_{k+2}=\ldots t_n-t_{n+1}=0$$
$$t_n-t_1=t.$$
(up to an irrelevant shift this is of the form
$(0,\ldots,t,\ldots,t)$
with $k$ zeroes and $n-k+1$ $t$'s).
After this shift, the equation \anc\ becomes
\eqn\deanc{xy+z^k(z+t)^{n-k+1}=0,}
which for $t\neq 0$ visibly has an $A_{k-1}$ singularity at $(0,0,0)$
and
an $A_{n-k}$ singularity at $(0,0,-t)$.
The vanishing cycles over $t$ are now given by the
 curves $C_1,\ldots,C_{k-1},C_{k+1},\ldots,C_n$ as $t$ varies.  These
naturally decompose into two connected components, with
$E_1,\ldots,E_{k-1}$ in the first component corresponding
to the $SU(k)$ gauge symmetry
and $E_{k+1},\ldots,E_n$
in the second component corresponding to $SU(n-k+1)$ gauge symmetry.
Note that by a change of coordinates, defining $z'=z+t$ the above
singularity takes the form $xy+z^kz'^{n-k+1}=0$ which was
considered in \sau\ where it was noted that based
on the D-brane analysis of \ref\bsv{M.~Bershadsky, V.~Sadov, and
C.~Vafa,
{\it D-strings on D-manifolds\/}, HUTP-95/A035, hep-th/9510225.}\
(applied to the intersecting 7-branes in this case)
one expects matter in $({\bf k},{\bf n-k+1})$, in agreement
with the above result.

\subsec{$D_n\to D_{n-1},A_{n-1},D_{n-r}\times A_{r-1}$}
We will consider three cases corresponding to the breaking
patterns:
$$i)\ \ SO(2n-2)\times SO(2)\subset SO(2n)$$
$$ii)\ \ SU(n)\times U(1)\subset SO(2n)$$
$$iii)\ \ SO(2n-2r)\times SU(r)\times U(1)\subset SO(2n)$$
In case $i)$ by the decomposition of adjoint
we expect one hypermultiplet in the fundamental ${\bf 2n-2}$
of $SO(2n-2)$.  In the case $ii)$ we expect one
matter hypermultiplet in the antisymmetric tensor
representation ${\bf n(n-1)/2}$ of $SU(n)$. In the case $iii)$
we expect one hypermultiplet in the $({\bf 2n-2r},{\bf r})$ of
$SO(2n-2r)\times SU(r)$ as well as one hypermultiplet in the
$({\bf 1,r(r-1)/2})$.

The first breaking pattern $i)$ corresponds to
choosing $(t_1,...,t_n)=(t,0^{n-1})$.
Plugging this into equation \dcar\ we find
$$xy+y^2z-z^{n-1}-t^2z^{n-2}=0.$$
We see the $D_n$ for $t=0$ but a $D_{n-1}$ for $t\ne0$.

The symmetry breaking $ii)$ corresponds to choosing $(t_1,...,t_n)=
(t,t,...,t)$.
Plugging this into equation \dcar\ gives
$$x^2+y^2z-{(z+t^2)^n-t^{2n}\over z}+2t^ny=0,$$
which is $D_n$ for $t=0$ and $A_{n-1}$ for $t\ne0$ as we now argue.
There is a singularity at $(0,t^{n-2},-t^2)$.  Accordingly, we note
that
near $z=-t^2$, this equation has leading behavior
$$x^2-t^2(y-t^{n-2}z)^2+{(z+t^2)^n\over t^2}=0,$$
which shows that we have an $A_{n-1}$ singularity for $t\ne0$.

Note that in \sau\ it was found that, unlike the $A_n$ case, one does
not seem to get intersecting $D_n$ singularities corresponding
to mixed matter representations.  This is actually explained
in our case by noting that there is no $U(1)$ which breaks
the $D_n$ to the product of two $D$'s.  The closest that we
come to a product structure is the case $iii)$ above, where
we take
$(t_1,\ldots,t_r,t_{r+1},\ldots,t_n)=(t,\ldots,t,0,\ldots,0)$.
Plugging this into \dcar\ gives
$$x^2+y^2z-z^{n-r-1}(z+t^2)^r=0.$$
Note that this is visibly a $D_n$ for $t=0$; for $t\not=0$
we have a product singularity, where at $(x,y,z)=(0,0,0)$
we have $D_{n-r}$ singularity and at $(x,y,z)=(0,0,-t^2)$ we have an
$A_{r-1}$ singularity.

\subsec{ $E_n$ cases}
In this section we consider some breakings starting
from the exceptional groups $E_6,E_7$ and $E_8$.  These
are somewhat more challenging because as mentioned before
the Cartan deformations of the singularity are more complicated.

\subsec{$E_6\to D_5,A_5$}
We consider two examples for the $E_6$ case corresponding
to the decomposition
$$i)\ \  SO(10)\times U(1)\subset E_6$$
$$ii)\ \  SU(6)\times U(1)\subset E_6$$
In the case $i)$ by the adjoint decomposition we expect a localized
matter in the ${\bf 16}$ of $SO(10)$, and for case $ii)$
we expect matter in the ${\bf 20}$ of $SU(6)$ (third rank
anti-symmetric tensor representation).

The case $i)$ occurs when
$(t_1,\ldots,t_6)=(t,-2t,t,t,t,t)$, from which the equation for the
singularity can be derived to be
%
$$-{x}^{2}+{{{z}^{4}}\over{4}}+{y}^{3}-3\,{t}^{2}y{z}^{2}-12\,{t}^{5}yz
-6\,{z}^{2}{t}^{6}-12\,{t}^{8}y-16\,{t}^{9}z-12\,{t}^{12}=0.$$
Shifting the singularity at $(0,0,-2t^3)$ to the origin, the equation
becomes
%
$$-{x}^{2}+{{{z}^{4}}\over{4}}-2\,{z}^{3}{t}^{3}+{y}^{3}-3\,{t}^{2}y{z}
^{2}=$$
$$-x^2-\left (2\,zt-y\right )\left (zt+y\right )^{2}+{z^4\over
4}=0,$$
which is visibly a $D_5$ singularity for $t\neq0$.

The second case occurs by choosing the $U(1)$ to correspond to
$(t_1,\ldots,t_6)=(-t/2,-t/2,-t/2,t,t,t)$, from which the
equation for the singularity
can be derived to be
%
$$-{x}^{2}+{{{z}^{4}}\over{4}}+{y}^{3}-{{9\,{t}^{2}y{z}^{2}}\over{8}}+
{{243\,{z}^{2}{t}^{6}}\over{512}}-{{2187\,{t}^{8}y}\over{4096}}+{
{19683\,{t}^{12}}\over{131072}}=0.$$
Translating the singularity at $(0,27t^4/64,0)$ to the origin, we get
$$-{x}^{2}+{{{z}^{4}}\over{4}}+{y}^{3}+{{81\,{y}^{2}{t}^{4}}\over{64}}
-{{9\,{t}^{2}y{z}^{2}}\over{8}}=0.$$
The terms ${{{z}^{4}}\over{4}}-{{9\,{t}^{2}y{z}^{2}}\over{8}}
+{{81\,{y}^{2}{t}^{4}}\over{64}}$ are a perfect square; the natural
change of coordinates reveals that there is an $A_5$ singularity for
$t\neq0$.

\subsec{$E_7\to D_6,E_6,A_6$}
We will consider three cases corresponding to the breaking patterns
$$i)\ \ SO(12)\times U(1) \subset E_7$$
$$ii)\ \ E_6\times U(1) \subset E_7$$
$$iii)\ \ SU(7)\times U(1)\subset E_7$$
  The respective representations are expected
to be one hypermultiplets in $SO(12), {\bf 32}$;
$E_6, {\bf 27}$; $SU(7), {\bf 35}\oplus {\bf 7}$.

The first case occurs by choosing the $U(1)$ to correspond to
$(t_1,\ldots,t_7)=(t,-2t,t,t,t,t,t)$, from which the
equation of the singularity can be derived to be
$$-{x}^{2}-{y}^{3}+16\,y{z}^{3}+36\,{t}^{2}{y}^{2}z=0.$$
The coordinate change $z\mapsto z+y/(36t^2)$ yields for $t\neq0$
$${2916\,{x}^{2}{t}^{6}-46656\,y{z}^{3}{t}^{6}-3888\,{z}^{2}{y}^
{2}{t}^{4}-108\,z{y}^{3}{t}^{2}-{y}^{4}-104976\,{t}^{8}{y}^{2}z}=0,$$
which is seen to be a $D_6$ after completing the square in $y$.

The second case occurs
by choosing $(t_1,\ldots,t_7)=(0,0,0,0,0,t,0)$, from which the
equation for the singularity can be derived to be
$$-{x}^{2}-{y}^{3}+16\,y{z}^{3}+3\,{t}^{2}{y}^{2}z+{{3\,{t}^{6}{y}
^{2}}\over{8}}-{{9\,{t}^{8}yz}\over{16}}+{{3\,{t}^{10}{z}^{2}}\over{16}} 
-
{{11\,{t}^{12}y}\over{256}}+{{9\,{t}^{14}z}\over{256}}+{{7\,{t
}^{18}}\over{4096}}=0.$$
Translating the singularity at $(0,t^6/16,-t^4/16)$ to the origin,
we get
%
$$3\,{t}^{2}{y}^{2}z-{x}^{2}-{y}^{3}-3\,y{z}^{2}{t}^{4}+16\,y{z}^{3}+{t} 
^{6}{z}^{3}=0.$$
The change of variables $y\mapsto y+t^2z$ changes this to
$$-{x}^{2}-{y}^{3}+16\,y{z}^{3}+16\,{z}^{4}{t}^{2}=0,$$
which visibly is an $E_6$ singularity for $t\neq0$.

The third case occurs by choosing
$(t_1,\ldots,t_7)=(-t,-t,-t,2t,2t,2t,2t)$,
from which the equation of the singularity can be derived to be
$$-{x}^{2}-{y}^{3}+16\,y{z}^{3}+63\,{t}^{2}{y}^{2}z+{{5103\,{t}^{6
}{y}^{2}}\over{8}}+{{45927\,{t}^{8}yz}\over{16}}+{{413343\,{t}^{10}{
z}^{2}}\over{16}}+$$
$${{40920957\,{t}^{12}y}\over{256}}-
{{90876411\,{t}^
{14}z}\over{256}}+{{35255264499\,{t}^{18}}\over{4096}}=0.$$
Translating the singularity at $(0,-2187t^6/16,-81t^4/16)$ to the
origin,
we get
$$63\,{t}^{2}{y}^{2}z-{x}^{2}-{y}^{3}-243\,y{z}^{2}{t}^{4}+16\,y{z}^{3}- 
2187\,{t}^{6}{z}^{3}+59049\,{t}^{10}{z}^{2}+$$
$$729\,{t}^{6}{y}^{2}-13122
\,{t}^{8}yz=0.$$
The last three terms form a perfect square of a linear expression in
$y,z$.  We adapt coordinates to this linear term to see that the
singularity
is $A_r$ for some $r$.  To find the
value of $r$, we make a linear coordinate change takes the equation
to the form
form $x^2+y^2+f(y,z)=0$ where $f$ has no terms of degree 2 or less.
The problem is that $f(y,z)$ contains a term $y^2z$ which must be
eliminated.
We make a change of variables
$z\mapsto z+\sum_{i=2}^k c_iy^i$ for $k=2,3,\ldots$, and recursively
solve for the $c_i$ in order to make the coefficient of $y^rz$ equal
to
0.  This happens first when $r=7$, and then a $y^8$ term appears,
showing that we have an $A_7$ singularity for $t\neq0$.

\subsec{$E_8\to E_7$}
We consider the decomposition $E_7\times U(1)\subset E_8$.
We expect matter in the
${\bf 56}$ of $E_7$.

This corresponds to choosing the $U(1)$ to be
$(t_1,\ldots,t_8)=(t,0,\ldots,0)$.   Following \km, Appendix~0,
the singularity can be described implicitly in projective coordinates
$(W,X,Y,Z)$ by the parameterization
$$W={x}^{3}-y{z}^{2},$$
$$X={y}^{8}z-3\,{t}^{3}{z}^{2}{y}^{7}-3\,{x}^{2}t{y}^{7}+2\,{t}^{3}{y}^{ 
6}
{x}^{3}+{t}^{6}{z}^{3}{y}^{6}-{t}^{6}z{y}^{5}{x}^{3}+3\,{t}^{2}z{y}^{7
}x,$$
$$Y=-{t}^{4}{x}^{3}{y}^{3}+{t}^{4}{y}^{4}{z}^{2}+x{y}^{5}-2\,t{y}^{5}z+{ 
t}
^{2}{x}^{2}{y}^{4}$$
$${Z}={y}^{3}-{t}^{2}x{y}^{2}.$$
These satisfy the equation
$$-{X}^{2}+{Y}^{3}-{Z}^{5}W+{t}^{2}Y{Z}^{3}W=0.$$
To get the affine equation of the singularity, we just put $W=1$;
this is
visibly an $E_7$ when $t\neq 0$.
The coefficients of this polynomial depend on $t^2$ rather than just
$t$.
In this case, the reason is that the ${\bf 56}$ denotes two
$1/2$-hypermultiplets.

\newsec{Matter from the geometry of the vanishing 2-cycles}

In this section, we note how the matter representations can also be
inferred
from the geometry of the vanishing 2-cycles.  We illustrate with a
single
example, although the method works for any configuration of vanishing
2-cycles corresponding to an A-D-E singularity.

Starting with an embedded $U(1)$, the positive roots which are
neutral with respect to the $U(1)$ correspond to vanishing 2-cycles
for
$t\ne0$.  For $t=0$, we get more vanishing 2-cycles.  We can reverse
this
reasoning to start with the geometry of the vanishing 2-cycles and
infer
the $U(1)$, hence infer the matter content.

As our example, in \ref\bkkm{P.~Berglund, S.~Katz, A.~Klemm, P.~Mayr,
{\it New Higgs Transitions between Dual $N=2$ String Models\/},
NSF-ITP-95-162, hep-th/9605154.}, it was noted that the Calabi-Yau
threefold
$X_5$ arising from resolving singularities of ${\bf
P}(1,1,2,6,8)[18]$
has families of vanishing 2-cycles which sweep out two divisors
(called
$D_8,D_9$ in~\bkkm).  For special parameter values, one of the
vanishing
2-cycles split into a pair of vanishing 2-cycles.

Choosing $t$ to be a parameter for the vanishing 2-cycles with $t=0$
at a
special parameter value, we name the vanishing 2-cycles at $t=0$ as
$C_1,C_2,C_3$, with $C_1$ lying in $D_8$ and $C_2,C_3$ lying in
$D_9$.
To yield the desired neutrality of $C_1$ and $C_2+C_3$, we take the
$U(1)$ in the direction given by $t_1=t_2=t_4$ in the $A_3$ Cartan.
This gives $SU(3)\times U(1)\subset SU(4)$, and we predict the ${\bf
3}$
of $SU(3)$, as was noted in the context of type IIA compactifications
in~\bkkm\ based on~\bsv.

We would like to thank James Cogdell, Michael Bershadsky,
Ken Intriligator, Shamit Kachru, Vladimir Sadov and
David Morrison for helpful
conversations.   C.V. would also like to thank
the hospitality of theory division at CERN where
this work was completed.
 The research of S.K.\ was supported in part by NSF
grant DMS-9311386 and NSA grant MDA904-96-1-0021.
The research of C.V.\ was supported in part by
NSF grant PHY-92-18167.

\listrefs
\bye